\documentstyle[12pt,aasms4,psfig]{article}
 
\slugcomment{to appear in the Astrophysical Journal} 
 
\begin{document}
 
\title{The X-ray -- radio alignment in the $z = 2.2$ radio galaxy
PKS~1138--262}
 
\author{C.L. Carilli}
\affil{National Radio Astronomy Observatory, P.O. Box O, Socorro, NM,
87801 \\ 
ccarilli@nrao.edu}
\author{D.E. Harris}
\affil{Smithsonian Astronomical Observatory, 60 Garden St., Cambridge,
MA, 02138 \\} 
\author{L. Pentericci}
\affil{Max-Planck Institute for Astronomy, Heidelberg, Germany \\} 
\author{H.J.A. R\"ottgering, G.K. Miley, J.D. Kurk}
\affil{Leiden Observatory, Postbus 9513, 2300 RA Leiden, The
Netherlands \\}  
\author{Wil van Breugel}
\affil{IGPP, Lawrence Livermore National Laboratories, Livermore, CA,
USA \\}

\begin{abstract}

We present high resolution X-ray observations of the narrow line radio
galaxy PKS 1138$-$262 at z = 2.156 with the ACIS-S detector on the Chandra
observatory. These observations show that the X-ray emission
from 1138--262 is dominated by emission from the active galactic
nucleus (AGN) with a 2 to 10 keV luminosity of $4\times10^{45}$ erg
s$^{-1}$. The relative X-ray and radio properties of the AGN in 
1138--262 are similar to those seen for
the AGN in the archetype powerful radio galaxy Cygnus A.

Between 10$\%$ and 25$\%$ (depending on energy)
of the X-ray emission from 1138--262 is
spatially extended on  scales of 10$''$ to 20$''$. The extended X-ray
emission is elongated, with a major axis aligned with
that of the radio source.  While the
X-ray and radio emissions are elongated on similar scales and position
angles, there is no one-to-one correspondence between the radio and
X-ray features in the source. The most likely origin for the
extended X-ray emission in 1138--262 is thermal emission from shocked
gas, although we cannot rule-out a contribution from  inverse Compton
emission. If the emission is from hot gas, the 
gas density is 0.05 cm$^{-3}$ and the gas mass is
$2.5\times10^{12}$ M$_\odot$.  The pressure in this hot gas is adequate
to confine the radio emitting plasma and the optical line emitting
gas.  We set an upper limit of $1.5\times10^{44}$ erg s$^{-1}$
to the (rest frame) 2 to 10 keV luminosity of any normal cluster
atmosphere associated with 1138--262.

No emission was detected from any
of the Ly$\alpha$ emitting galaxies in the (proto-) cluster around
1138--262, outside of the Ly$\alpha$ halo of 1138--262 itself,
to  a (rest frame)  2 to 10 keV luminosity limit of $1.2\times10^{43}$
erg s$^{-1}$. Emission was detected from a $ z = 2.183$ QSO located
2$'$ west of 1138--262 with a luminosity of $1.8\times10^{44}$ erg
s$^{-1}$.

\end{abstract}

{\sl Subject headings:} cosmology -- large scale structure; galaxies
-- active; radio continuum -- galaxies; X-rays -- galaxies, clusters)

\section {Introduction}

Radio galaxies are the most massive galaxies known at
high redshift, and have been shown to reside in regions of high
ambient gas and galaxy density (R\"ottgering, Best, \& Lehnert 1999;
McCarthy 1993).  
As such,  these sources provide unique laboratories for the study of 
the formation of galaxies and clusters of galaxies,
acting as beacons to biased galaxy formation within large scale
structure at high $z$ (Carilli et al. 2001). 

One of the more extensively studied high redshift
radio galaxies in a dense environment is the source 
PKS 1138--262 at $z=2.156$ (Pentericci et
al. 1997). The 1138--262 radio source has the most disturbed
morphology of any $z > 2$ radio galaxy yet identified, appearing as a
string of bright knots with a total linear extent of 15$''$ (Carilli
et al. 1997). The radio source is enveloped in a Ly$\alpha$ emitting
halo with a luminosity of $2.5\times 10^{44}$ erg s$^{-1}$, spatially 
extended on a scale of 20$''$ and with a 
major axis oriented along the radio source axis (Pentericci et
al. 1997). There are clear spectroscopic signatures of strong
interaction between the expanding radio source and the ambient gas
(Kurk et al. 2001).  The polarized radio emission from 1138--262 shows
the most extreme values of Faraday rotation of any source at $z > 2$,
with (rest frame) rotation measures (RMs) up to 6250 rad m$^{-1}$
(Carilli et al. 1997).  These RMs are comparable to values observed
toward low $z$ powerful radio galaxies in dense, X-ray emitting cluster
atmospheres.  For the low $z$ sources the origin of the large RMs is
thought to be magnetized cluster gas, with magnetic fields 
of a few $\mu$G or more (Carilli \& Taylor 2001).

High resolution optical images show that the parent galaxy of 
1138--262 is comprised of many compact `knots' of emission distributed
on a scale of 10$''$. Wide field imaging and spectroscopy with the VLT
has revealed many compact Ly$\alpha$ emitting galaxies distributed on
arcminute scales around 1138--262 (Pentericci et al. 2000), with an
implied over-density of 
objects similar to $z \sim 3$ proto-clusters recently discovered in
searches for UV dropout galaxies by Steidel et al. (1998).  Pentericci
et al. (2000) interpret these results on 1138--262 as indicating a
forming giant elliptical galaxy within a proto-cluster.

Carilli et al. (1998) searched for X-ray emission from 1138--262
using the HRI detector on ROSAT. While the source was clearly
detected, the sensitivity and spatial resolution of the ROSAT
observations were insufficient to determine the relative contribution 
to the X-ray luminosity from the AGN and possible extended emission.
In this paper we present Chandra observations which
clarify the X-ray emission processes in  1138--262. While
the dominant X-ray source is the AGN, we also detect an extended 
soft  X-ray emission component with an extent and orientation 
similar to the radio source. We consider mechanisms for this 
extended emission, and discuss the origin of the X-ray -- radio
alignment in 1138--262. We use H$_o$ = 65 km s$^{-1}$ Mpc$^{-1}$,
$\Omega_M = 0.3$, and $\Omega_\Lambda = 0.7$, leading to
D$_{\rm L}$ = 18300 Mpc and D$_\theta$ = 1844 Mpc (1$''$ = 9 kpc).
We defined spectral index, $\alpha$, as a function of frequency, 
$\nu$,  and flux density, S$_\nu$, as: $\rm S_\nu \propto
\nu^{\alpha}$. 

\section{Observations and Results}

PKS 1138--262 was observed for 40 ksec on June 6, 2000, 
with the back-illumiated
ACIS-S CCD detector array on the Chandra Observatory (Weisskopf,
O'Dell, \& Van Speybroeck,  1996; Garmire 1997). The source was
positioned one arcmin from the  standard aim point 
so as to avoid the node boundary on the ACIS-S3 chip.
Standard ACIS settings were used: TE mode with 3.2 s readout and 'faint' 
telemetry format. The data were reprocessed in 2001Jan.
For analysis of extended regions the event file was filtered to reject 
times of high background, resulting in a 
a live-time of 28 ksec.  For unresolved sources we use the entire observation 
(live-time of 39.5 ksec).   Data analysis 
was performed using the CIAO 2.1 software package for
spectroscopic imaging. Image analysis was also performed using
the Astronomical Image Processing System. 

\subsection{Images}

Figure 1 shows the Chandra total X-ray image (energy range = 0.3 to 10
keV), convolved with a Gaussian of Full Width at Half Maximum (FWHM)
of 2$''$, along with the VLA radio image at 1.4 GHz, 2$''$ resolution
in greyscale (from Carilli et al. 1997). 
A total of 880 counts are detected in a $17'' \times
22''$ box centered on the radio galaxy. We estimate a background
contribution in this box of 102 counts.  

Two components are evident in Figure 1: (i)  compact emission 
at the location of the AGN, and (ii) extended emission on a scale
of about 20$''$.  
Comparison of the compact AGN component to the Chandra point spread
function, and Gaussian fitting, shows that the emission is unresolved
with a size $< 1''$.  The total emission from the nucleus is 663 counts.
There is also a point source about 5$''$ northwest of the nucleus
which contains 15 counts at a position of   11$^h$ 40$^m$ 47.93$^s$,
--26$^o$ 29$'$ 06.5$''$ (J2000). This source is also detected in the
Ly$\alpha$  and the HST images, but not in the 5 GHz image
to a 3$\sigma$ limit of 90$\mu$Jy (Figure 2a). 
Subtracting the background and the two point sources
leads to 98$\pm$30 counts for the extended emission from 1138--262, or
13$\%$ of the total.

Figure 2a shows the image of total X-ray emission from 1138--262
convolved with a Gaussian of FWHM = 1$''$, along with the radio image
at 5GHz, 0.5$''$ resolution (2a), the HST F606W image (2b), and the
VLT Ly$\alpha$ image (2c). The position of the nucleus of 1138--262 as
determined in near IR images, H$\alpha$ line images, and the HST
image, is 11$^h$ 40$^m$ 48.37$^s$, --26$^o$ 29$'$ 08.9$''$ $\pm$
0.1$''$ (J2000) using USNO astrometry (Monet et al. 1996).  The radio
nucleus is 
located at 11$^h$ 40$^m$ 48.35$^s$, --26$^o$ 29$'$ 08.8$''$ $\pm$0.1$''$,
implying that the radio and optically determined positions for the
nucleus are within 0.2$''$ of each other.  The nominal astrometry from
Chandra places the X-ray nucleus at 11$^h$ 40$^m$ 48.31$^s$, --26$^o$
29$'$ 08.6$''$, or $-0.5\pm0.1''$ and $-0.2\pm0.1''$ away from the
radio position in RA and Dec, respectively. This is consistent with
the offset seen for the $z = 2.183$ QSO in the 1138--262 field
(see below), and with the mean offsets of $-0.4''$ and $0.0''$ in RA
and Dec, respectively, between radio and X-ray positions of sources in
the Abell 370 field found by Barger et al. (2001).  In the
analysis below, the images at different frequencies are aligned using
the nucleus.

Figure 3 shows the X-ray emission in three energy bands: 0.3 to 1.2 keV
(3a = soft), 1.2 to 2.4 keV (3b = mid), and $> 2.4$ keV (3c = hard). 
The 5 GHz radio continuum
emission is shown in greyscale in Figure 3a. 
The extended X-ray emission is softer 
than the emission from the nucleus, while the emission from the point
source 5$''$ northwest of the nucleus is harder. For the soft
image the extended emission contributes $21\pm 3.5 \%$ to the total
source counts. This percentage drops to $7\pm 1\%$ for the mid energy
range, and $< 11\%$ in the hard X-ray image.

Figure 4 shows the surface brightness profiles averaged azimuthally in
rings centered on the nucleus of 1138--262 for quadrants oriented
parallel and perpendicular to the radio axis.  The profile along the
radio axis shows extended emission to about 12$''$ radius.  For
illustrative purposes we have fit a standard cluster $\beta$ model to
the on-axis profile, as shown with the solid line. The formal results
for the model parameters are $\beta = 2.5\pm 1$ and $\rm r_c =
12\pm4''$. This value of $\beta$ is much steeper than what is expected
for a typical cluster atmosphere, even for high redshift clusters, for
which $\beta \sim {2 \over 3}$ is observed typically (Stanford et
al. 2001).  Such a steep $\beta$ may simply reflect the fact that the
extended emission region is clearly not circularly symmetric, but is
highly elongated. An elongated morphology will lead to a decrease in
the fractional emitting area intercepted with increasing annular
radius, and hence an artificially steep radial profile. Perhaps a more
interesting result in Figure 4 is that the profile perpendicular to
the radio source axis shows very little emission beyond about 4$''$
distance from the nucleus.  Averaging the signal out to 12$''$ radius,
excluding the inner 5$''$, shows that the luminosity of the extended
emission perpendicular to the radio source major axis is less than
25$\%$ of that seen along this axis.

We have searched for X-ray emission from the
Ly$\alpha$ emitting galaxies in the
(proto) cluster around 1138--262 (Pentericci et al. 2000).  There are no
detected X-ray sources within 1$''$ of any of the cluster galaxies,
outside of the Ly$\alpha$ halo of 1138--262 itself.  The 3$\sigma$
upper limit to the 2 to 10 keV X-ray luminosities of these sources is
$1.2\times10^{43}$ erg s$^{-1}$, assuming a standard powerlaw spectrum
of spectral index $\alpha = -0.8$. X-rays are detected from a $z =
2.183$ QSO located 2$'$ west of 1138--262 (Pentericci et al. 2000), at
11$^h$ 40$^m$ 39.73$^s$ --26$^o$ 28$'$ 44.9$''$ (J2000). A total of 56
counts are detected from this source, implying an X-ray luminosity of
$1.8 \times10^{44}$ erg s$^{-1}$ for $\alpha = -0.8$.

\subsection{The X-ray Spectra}

The X-ray spectrum of the nuclear emission from 1138--262 is shown in
Figure 5. The spectrum is reasonably fit by a power-law with $\alpha =
-0.8\pm0.2$, plus absorption at the redshift of the source of N(HI) =
$2.6 \pm 0.6 \times 10^{22}$ cm$^{-2}$.  The reduced $\chi^2$ for the
fit is 0.52, and the error bars for the fit parameters represent
changes in reduced $\chi^2$ by $\pm 1$.  The Galactic value of N(HI)
in this direction is $4.5\times 10^{20}$ cm$^{-2}$ (Stark et
al. 1992). The unobscured 2 to 10 keV luminosity for the nucleus is
$\rm L_{X}^{nuc} = 4.0 \times 10^{45}$ erg s$^{-1}$.  The implied
extinction toward the nucleus is $\rm A_V = 16$, assuming a Galactic
dust-to-gas ratio.  This large extinction may explain the lack of strong
Ly$\alpha$ emission from the nucleus of 1138--262. For completeness,
we note that the nuclear spectrum can be equally well fit
by a thermal  spectral model (Raymond-Smith) with $\rm kT =
16_{-4}^{+7}$ keV. 

Considering the radio properties of 1138--262, 
the luminosity between rest frame frequencies of
1 to 10 GHz for the  nucleus of  1138--262  is: $\rm  L_{1-10
GHz}^{nuc} = 1.5\times10^{43}$ erg s$^{-1}$.
The radio luminosity for the entire source
for rest frame frequencues of 0.1 to 1.0 GHz is:
$\rm  L_{0.1-1 GHz}^{tot} = 1.8\times10^{45}$ erg s$^{-1}$
(Carilli et al. 1997). 

The archetype powerful radio galaxy
Cygnus A has a highly absorbed X-ray emitting nucleus with
N(HI) =  $3.8\times 10^{23}$ cm$^{-2}$ and 
an unobscured 2 to 10 keV luminosity of  $\rm L_X^{nuc} = 1\times
10^{45}$ erg s$^{-1}$ (Ueno et al. 1994). The radio nucleus of Cygnus
A has  $\rm  L_{1-10 GHz}^{nuc} = 1.3\times10^{42}$ erg s$^{-1}$, 
and the radio  luminosity for the entire source from 0.1 to 1.0 GHz 
is  $\rm  L_{0.1-1 GHz}^{tot} = 3.8\times10^{44}$ erg s$^{-1}$
(Carilli et al. 1991). Hence, to within a factor two for both 
1138--262 and Cygnus A:  
$\rm {{L_{X}^{nuc}}\over{L_{1-10 GHz}^{nuc}}} \sim 500$,
and $\rm {{L_{X}^{nuc}}\over{L_{0.1-1 GHz}^{tot}}} \sim 2.5$.
Ward et al. (1996) show that the obscuration corrected
properties of the X-ray nucleus of Cygnus A 
are typical of broad line radio galaxies.

Due to the paucity of counts, the spectrum of the extended X-ray
emission from 1138--262 is
very poorly constrained, with an allowed range 0.8 to 15 keV for a
Raymond-Smith thermal plasma emission model.  In the analysis below we
adopt a value of 5 keV.  The implied 2 to 10 keV (rest frame)
luminosity for the extended emission is then $3\times 10^{44}$ erg
s$^{-1}$.  Further observations of 1138--262 are required to constrain
this important parameter.

\section{Discussion}

The most interesting aspect of the extended X-ray emission 
from 1138--262 is that it
is oriented along the radio axis and has a similar spatial extent.
There are a number of possible mechanisms to explain this emission,
and its alignment with the radio axis.

\subsection{Inverse Compton, synchrotron, and scattering}

One mechanism for the extended X-ray emission from 1138--262 
is synchrotron radiation from the relativistic electrons
in the radio lobes. The total extended X-ray emission
corresponds to $1\times10^{-8}$ Jy at 2 keV. The radio spectral index
for the integrated emission from  1138--262 at cm wavelengths is
--1.1, and  the integrated flux density from the source at 5 GHz is
160 mJy (Carilli et al. 1997). Extrapolating the radio spectrum 
to  2 keV ($4.8\times 10^{17}$ Hz) implies an expected flux density
of $3\times 10^{-10}$ Jy. Hence the X-ray emission cannot be a
simple extrapolation of the radio synchrotron spectrum.
The spectrum  would have to flatten at frequencies above 15 GHz.
Such a flattening can be 
ruled-out by the fact that the predicted I-band ($3.7\times10^{14}$
Hz) emission would be $6 \times 10^{-6}$ Jy, while the upper limit 
is $1 \times 10^{-8}$ Jy. 


A second possibility is Thompson scattering of X-ray emission from
the nucleus. This is easily ruled out since the spectrum of the
extended  X-ray emission is different than that of the nucleus.

A third possible mechanism is inverse Compton (IC) emission, ie.
up-scattering of the cosmic microwave background (CMB) by the
relativistic electrons in the radio source.  In this case comparing
the radio synchrotron and X-ray IC emission constrains the magnetic
fields within the radio source, since both depend on the relativistic
electron population.  Using equation 11 in Harris and Grindlay (1979),
and using the integrated flux densities of the extended emission from
1138--262 in the radio and X-ray, the resulting magnetic field is
7$\mu$G assuming $\alpha = -1$. For comparison, the minimum energy
magnetic field is about 200 $\mu$G in the radio source (Carilli et al.
1998). Whether the fields and particles 
conspire to reach a minimum configuration remains
unknown, however, a field value a factor 30 lower than the minimum
energy value raises the internal pressure by a factor 150.
If the fields are at the minimum energy level, then
IC emission constitutes 0.1$\%$ of the total extended X-ray emission. 

Brunetti et al. (2001) have shown that in some sources 
the energy density in the photon field  in the radio lobes 
due to  optical/IR photons from
the AGN is larger than that in the CMB. In principle, this would
allow  for larger magnetic fields in the radio lobes. 
However, assuming a luminous (obscured) AGN of optical through IR
luminosity $\sim 10^{46}$ erg s$^{-1}$ 
in 1138--262 (as adopted by Brunetti et al. for 3C 295), 
and correcting the energy density in the CMB for redshift, 
it is clear from their Figure 2 that the energy density
in the photon field will be dominated by the CMB 
beyond about 10 kpc from the nucleus. The 
X-ray emission in 1138--262 extends well beyond this radius.

Perhaps the most telling argument against IC being the
dominant emission mechanism for the X-rays seen from
1138--262 is that there are off-sets between
the positions of 
high surface brightness radio and X-ray emitting features
(see Figures 1 and 2). In all radio galaxies for which 
IC X-ray emission has been seen, it is coincident
with high surface brightness radio features (Harris et al. 
2000, Harris, Carilli, \& Perley 1994). If the dominant photon field
is the microwave background, the observed X-rays are predominantly
from electrons with Lorentz factors $\rm \gamma_{IC} = 1000$, 
corresponding to radio emission at
270 MHz for $\rm B = 200\mu G$. These data 
imply that if the X-ray emission is IC,
then the radio morphology of the source would have to
change significantly from low to high frequencies.
High resolution images at frequencies $\le 500$ MHz are
required to test this hypothesis, although 1.4 GHz images
show no dramatic change in the source morphology relative
to that seen at 5 GHz. 

\subsection{Hot Gas}

A final mechanism we consider is thermal emission from hot gas.  The
extended emission from 1138--262 has a 2 to 10 keV 
luminosity of $3 \times 10^{44}$
erg s$^{-1}$.  For comparison, the cluster atmosphere enveloping
the powerful radio galaxy  Cygnus A has a 2 to 10 keV luminosity
of $4 \times 10^{44}$ erg s$^{-1}$ (Arnuad et al. 1987). 
The Cygnus A cluster atmosphere is roughly azimuthally
symmetric, with a FWHM $\sim 400$ kpc (Reynolds and
Fabian 1996).  The spatial
extent of the X-ray emission from 1138--262 is confined to a region of
$180 \times 90$ kpc, and is aligned along the radio axis.  So while
the extended X-ray luminosities are within a factor two for Cygnus A
and 1138--262, the relative volumes differ by a factor of about
30. The implication is that the density of hot gas in  1138--262
must be a factor of five or so larger than that seen in Cygnus A,
implying gas densities $\sim 0.05$ cm$^{-3}$.  The mass in hot gas is
$2.5\times10^{12}$ M$_\odot$.

The pressure in the hot gas is $\sim 8\times10^{-10}$ dynes cm$^{-2}$,
assuming $\rm T = 6\times 10^7$ K.  The minimum pressure in the radio
source is $\sim 6\times10^{-10}$ dynes cm$^{-2}$, while the pressure in the
Ly$\alpha$ emitting gas is about $\sim 1\times10^{-9}$ dynes cm$^{-2}$
(Pentericci et al. 1997). The pressure in the hot gas appears
to be roughly adequate
to confine both the radio source and the low filling factor, line
emitting clouds. Note that this confinement mechanism applies only to
the Ly$\alpha$  emitting regions within the maximum radius of the
X-ray source.

The cooling time for such  gas is $1\times 10^9$ years,
which is more than an order of magnitude larger than the dynamical
timescale for  the system, but a factor three shorter than
the age of the universe at that redshift. 
A possible heating mechanism of this gas is 
the expanding radio source, ie. the X-ray emission is from the
`cocoon' of shocked material enveloping the radio source, as
predicted by dynamical models of supersonically expanding 
jets (Begelman, Blandford, \& Rees 1986).
Assuming that the radio luminosity
is $\le 10\%$ of the jet kinetic luminosity 
implies a lower limit to the jet kinetic luminosity of 
$5\times10^{46}$ ergs s$^{-1}$.  The total energy stored
in the X-ray gas is about $2\times 10^{61}$ ergs. So it appears
that the radio jet could have heated the medium over a 
timescale of $2\times10^7$ years, which is reasonable for 
a radio galaxy of the size of 1138--262 (Blundell \& Rawlings 1999). 

Our search for X-ray emission from  1138--262 was originally
motivated by the extreme values of Faraday rotation, with the idea
that the source is embedded in a dense, magnetized cluster
atmosphere. While we have detected extended X-ray emission, the
morphology of the emitting regions is distinctly different than
that expected for a normal cluster atmosphere.  So whence the large
rotation measures? If the X-ray emission is from the coccoon of
shocked gas around the radio lobe, then large rotation measures will
occur if the shocked material contains magnetic fields of order 20
$\mu$G, ordered on scales of a few kpc (Bicknell, Cameron,
\& Gingold 1990). 

The lack of extended emission perpendicular to the radio source axis
allows us to set an upper limit to the 2 to 10 keV luminosity of an
extended (relaxed) cluster atmosphere around  1138--262 of $1.5 \times
10^{44}$ erg s$^{-1}$, or $< 40\%$ of the X-ray luminosity of the
Cygnus A cluster.  So while the 
overdensity of galaxies in the region suggests that the system will
evolve into a massive cluster, the lack of an extended X-ray
atmosphere is consistent with the idea that it is not yet a
dynamically relaxed system. This is also suggested by the irregular
velocity distribution of the Ly$\alpha$ emitting galaxies 
(Pentericci et al. 2000).

The lack of a hot cluster atmosphere also implies that the
ambient medium into which the radio source is expanding in 
1138--262 is very different than the diffuse cluster atmospheres
at $\sim 10^8$ K  seen in similar luminosity 
radio galaxies at lower redshift, 
such as Cygnus A and 3C 295. 
This difference is consistent with the suggestion by
Rees (1989) that the environments of  $z > 2$ radio galaxies 
consist of a multiphase medium of cool ($\le 10^4$ K) clouds or
filaments, embedded in a virialized medium at 10$^6$ to 10$^7$ K. 
The passage of the  bow shock driven by 
the expanding radio source would have a number of effects
in such a multiphase medium (Kaiser \& Alexander 1999).
Rees (1989) points out that the densest clouds are relatively
unaffected by the passage of the shock, and may be induced
to collapse and form stars by the high pressure medium in which
they suddenly find themselves ({\sl cf.} Icke 1999). The intermediate
density clouds are shocked, but the cooling time is shorter than
the jet lifetime, perhaps giving rise to some of the 
Ly$\alpha$ emission directly associated with  bright radio features.
The lower density, higher filling factor gas is shock heated 
to $\sim 10^{7 - 8}$, giving rise to the  extended X-ray emission.   
 
Barthel and Arnaud (1996) have proposed a selection effect in radio
galaxies in which jets propagating through regions of higher gas
density have a higher conversion efficiency of jet kinetic energy into
radio luminosity (see also Eales 1992).  If the ambient medium is
anisotropically 
distributed, then radio jets of a given kinetic luminosity propagating
along the direction of highest ambient density will be brighter than
ones propagating through low density regions. In this case,
preferential alignment between the jet axis and the axis of highest
density of the ambient material will occur naturally in radio flux
limited samples (West 1999). 

A final comparison we make is with the recent observations of the
extended X-ray emission around the $z = 1.79$ radio galaxy 3C 294 by
Fabian et al. (2001). The observations of 3C 294 show a similar
alignment between the X-ray and radio axes as that seen in 1138--262,
and a similar lack of one-to-one correspondence between high surface
brightness radio and X-ray features. In particular, for 3C 294 the
X-ray emission extends significantly beyond the radio source maximum
radius.  Fabian et al. (2001) interpret the extended X-ray emission as
being thermal emission from a cooling flow cluster atmosphere. Using
the temperature-luminosity relation for X-ray emitting cluster
atmospheres, they show that the existence of such a system at this
high redshift argues for a low $\Omega_M$ universe.  Inherent in this
argument is that the gas temperature is a gross measure of the
gravitational mass of system. For 1138--262 we have argued that the
gas heating is due to the expanding radio source, and hence unrelated
to the gravitational potential of the system.

The existence of a hot, high filling factor, thermal gas in high $z$
radio galaxies has long been hypothesized  in order to confine  both the
radio source and the optical line emitting clouds (Chambers et
al. 1990, Bremer et al. 1992). 
It may be that in 1138--262 the Chandra observatory has revealed 
this pervasive  hot medium.

\vskip 0.2in

The National Radio Astronomy (NRAO) is a facility
of the National Science Foundation, operated under cooperative
agreement by Associated Universities, Inc..
DEH and CLC acknowledge support from NASA grant GO0-1137B.
DEH acknowledges support from NASA contract NAS8-39073.
We acknowledge support from a programme subsidy provided by the 
Dutch Organization for Scientific Research (NWO).

\newpage

\centerline{\bf References}

Arnaud, K.A., Fabian, A.C., Eales, S.A, et al. 1987, MNRAS, 227, 241

Barger, A.J., Cowie, L.L., Mushotzky, R.F., \& Richards, E.A. 2001,
AJ, 121, 622

Barthel, P.D. \& Arnaud, K. 1996,  M.N.R.A.S. (letters), 283, 45

Begelman, M.C., Blandford, R.D., \& Rees, M.J. 1984, {\sl Rev. Mod.
Phys.}, 56, 255.

Bicknell, G.V., Cameron, R.A., \& Gingold, R.A.\ 1990, \apj, 357, 373 

Blundell, K.M. \& Rawlings, S. 1999, Nature, 399, 330

Bremer, M. N., Fabian, A. C., Sargent, W. L. W., Steidel, C. C.,
Boksenberg, A., Johnstone, R. M. 1992, MNRAS, 258, 23L

Brunetti, G., Cappi,  M., Setti, G., Feretti, L., \& Harris, D.E.\
2001, \aap, 372, 755  

Carilli, C.L., Perley, R.A., Dreher, J.W., \& Leahy, J.P.\ 1991, \apj, 
383, 554 

Carilli, C.L., R\"ottgering, H.J.A., van Ojik, R., Miley, G.K., \&
van Breugel, W.J.M. 1997,  Ap.J. (Supp.), 109, 1

Carilli, C.L., Harris, D.E., Pentericci, L., Roetterging, H.J.A.,
Miley, G.K., \& Bremer, M.N. 1998, ApJ, 496, 57L

Carilli, C.L., Miley, G.K., R\"ottgering, H.J.A., et al. 
2001,  in {\sl Gas \& Galaxy Evolution}, 
(ASP: San Francisco), eds. J. Hibbard, M. Rupen, \& J. van Gorkom.

Chambers, K. C., Miley, G. K., van Breugel, W. J. M. 1990,
ApJ, 363, 21

Eales, S.A. 1992, ApJ, 397, 49

Fabian, A.C., Crawford, C.S., Ettori, S., \& Sanders, J.S. 2001, 
MNRAS, 322, L11

Garmire, G.P. 1997, BAAS, 190, 3404

Harris, D.E. \& Grindlay, J.E. 1979, MNRAS, 188, 25

Harris, D.E., Carilli, C.L., \& Perley, R.A. 1994, Nature, 367, 713

Harris, D.E., Nulsen, P.E., Ponman, T.J. et al. 2000, ApJ, 530, L81

Icke, V.\ 1999, The Most Distant Radio Galaxies, (Royal Netherlands
Acadamy: Amsterdam), eds. H. Roettgering, P. Best, \& M. Lehnert, 217 

Kaiser, C.R. \& Alexander, P. 1999, MNRAS, 305, 707

Kurk, J.D., Roettgering, H.J.A., Pentericci, L., Miley, G.K., \&
Pentericci, L. 2001, Rev Mex. A\& A, in press

McCarthy, P.J. 1993,  A.R.A.A., 31, 639

Monet, D., Bird, A., Canzian, B., et al.  1996, USNO-SA1.0, (U.S.
Naval Observatory, Washington DC)

Miley, G.K. 1980, A.R.A.A., 18, 165

Pentericci, L., R\"ottgering, H.J.A.,  Miley, G.K., Carilli, C.L., \&
McCarthy, P. 1997,  A\&A, 326, 580

Pentericci, L., R\"ottgering, H.J.A.,  Miley, G.K., et al.
1998,  ApJ, 504, 139

Pentericci, L., Kurk, J.D., R\"ottgering, H.J.A. et al. 
2000, A\& A, 361, L25

Rees, M.J. 1989, MNRAS, 239, 1P

Reynolds, C.S. \& Fabian, A.C 1996,  M.N.R.A.S., 278, 479

Roettgering, H.J.A., Best, P., \& Lehnert, M.  1999, {\sl 
The Most Distant Radio Galaxies}, (Royal Netherlands Acadamy:
Amsterdam)

Stanford, S.A., 
Holden, B., Rosati, P., Tozzi, P., Borgani, S., Eisenhardt, P.R., \& 
Spinrad, H.\ 2001, \apj, 552, 504 

Stark, A.A., Gammie, C.F., Wilson, R.W. et al. 
1992,  Ap.J. (Supp.), 79, 77

Steidel, C., Adelberger, K., Dickinson, M.,  Giavalisco, M.,
Pettin, M., \& Kellogg, M. 1998, ApJ, 492, 428

Taylor, G.B. \& Carilli, C.L. 2002, ARAA, in press for Vol 40

Ueno, Shiro, Koyama, Katsuji, Nishida, Minoru, Shigeo, Yamauchi, \&
Ward, Martin J. 1994,  Ap.J. (letters), 431, L1

Ward, M.J. 1996, in {\sl Cygnus A}, eds. C. Carilli \& D. Harris, 
(Cambridge Univ. Press: Cambridge), p.43

Weisskopf, M.C., O'Dell, S.L., \& Van Speybroeck, L.P. 1996,
Proc. SPIE, 2805, 2 

West, M.J. 1999, in {\sl The Most Distant Radio Galaxies},
(Royal Netherlands Acadamy: Amsterdam), eds. 
H. Roettgering, P. Best, and M. Lehnert p. 365

\clearpage
\newpage

\centerline{Figure Captions}

\noindent Figure 1 -- The contours show the total (0.3 to 10 keV) 
X-ray emission from the $z = 2.156$ radio galaxy
PKS 1138--262 as observed with the ACIS-S detector
on the Chandra Observatory. The Chandra image has been convolved with
a Gaussian of FWHM = 2$''$. In this and subsequent images
the contours are a geometric progression
in square root two. The first contour level is 2.3 counts per
Gaussian beam. The greyscale shows the radio continuum 
image  at 1.4 GHz, 2$''$ resolution made with the Very Large Array 
(Carilli et al. 1997). The greyscale range is 0 to 150 mJy
beam$^{-1}$. 


\noindent Figure 2 -- A comparison of the X-ray emission
and emission at other wavelengths in PKS 1138--262. The contours in
all cases are of total X-ray emission
convolved with a Gaussian of FWHM = 1$''$.
The first contour level is 0.9 counts per Gaussian beam
(note: the highest four contour levels in these images are
by a factor two, not root two). The crosses mark
the position of the radio galaxy nucleus, and of an X-ray point source
located 5$''$ northwest of the nucleus. \\ 
a. The greyscale shows the 5 GHz radio continuum emission at 0.5$''$ 
resolution. The greyscale range is 0 to 2 mJy beam$^{-1}$. 
(from Carilli et al. 1997). \\
b. The greyscale shows the HST F606W  image (from Pentericci et
al. 1998). Note that 606nm corresponds to 192nm at $z = 2.2$. \\
c. The greyscale shows the VLT image of the Ly$\alpha$ emission (from
Kurk et al. 2001).  

\noindent Figure 3 -- Three-band  images of X-ray emission 
from PKS 1138--262 convolved with a Gaussian of FWHM = 2$''$.
The crosses are the same as in Figure 2. \\
a. The soft X-ray (0.3 to 1.2 keV) emission. The greyscale shows
the radio continuum emission at 5 GHz. The first contour level is 1.1
counts per Gaussian beam. \\
b. The mid X-ray (1.2 to 2.4 keV) emission. The first contour level is
0.9 counts per Gaussian beam.  \\
c. The hard X-ray ($>$2.4 keV) emission. The first contour level is
1.8 counts per Gaussian beam.

\noindent Figure 4 -- The surface brightness profiles averaged
azimuthally in rings centered on the nucleus of 1138--262.
The solid circles are for the two quadrants along the radio axis,
while the open squares are for the two quadrants perpendicular to the
radio axis. Both profiles have been normalized at the first point,
corresponding to a radius of 3.5$''$, and the points at smaller radii
are omitted due to confusion by AGN emission. The solid line is
a $\beta$ model fit to the  data along the radio axis, with
$\beta = 2.5\pm 1$ and $\rm r_c = 12\pm 4''$. The dotted
line is the background level. 

\noindent Figure 5 -- The X-ray spectrum of the AGN emission from PKS
1138--262. The model 
is a power-law spectrum with $\alpha = -0.8$ and N(HI) = $3.5 \times
10^{22}$ cm$^{-2}$ at $z = 2.156$. \\

\clearpage
\newpage

\begin{figure}
\psfig{figure=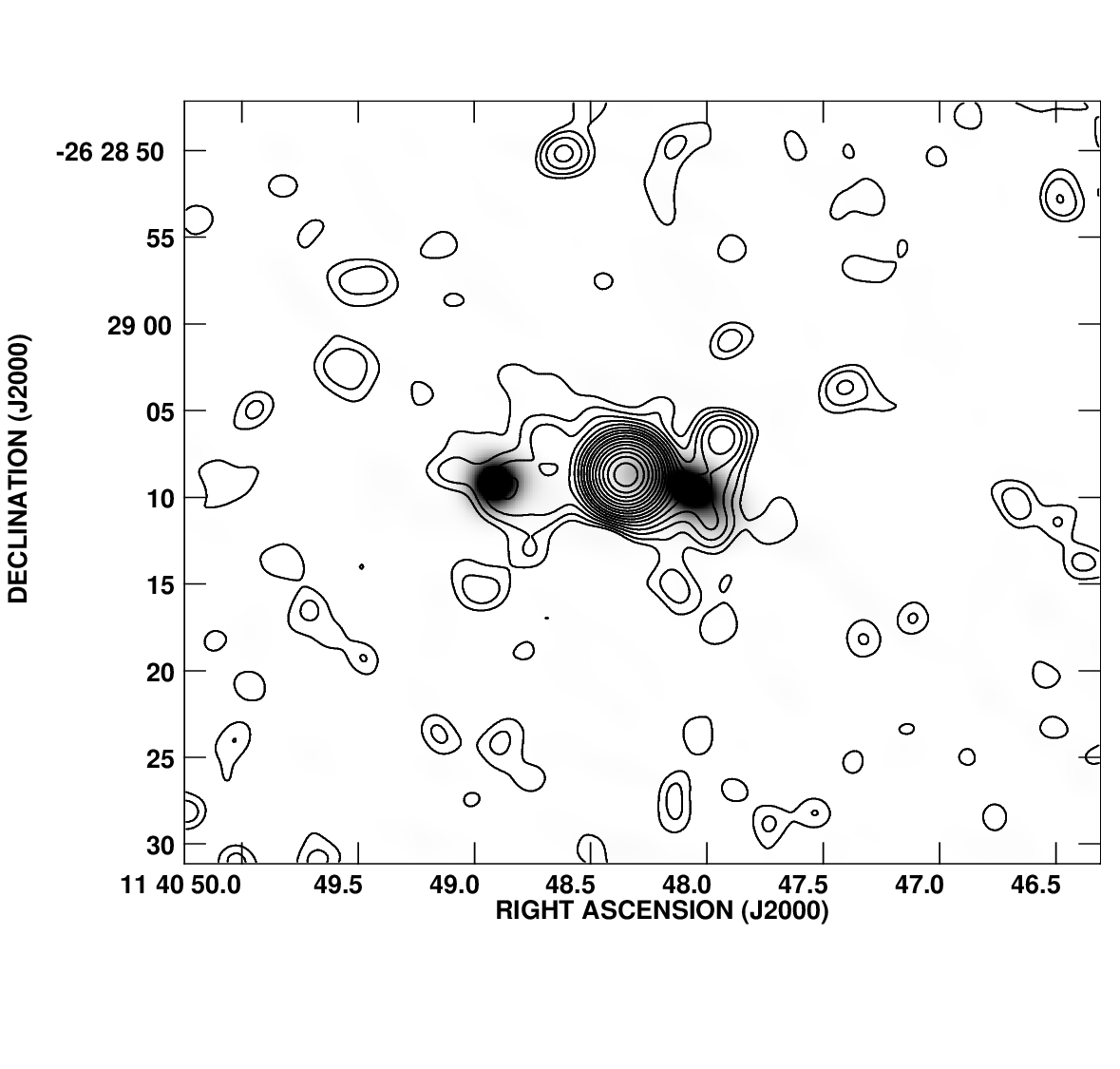,width=6in,angle=-90}
\caption{}
\end{figure}

\clearpage
\newpage

\begin{figure}
\vskip -0.5in
\psfig{figure= 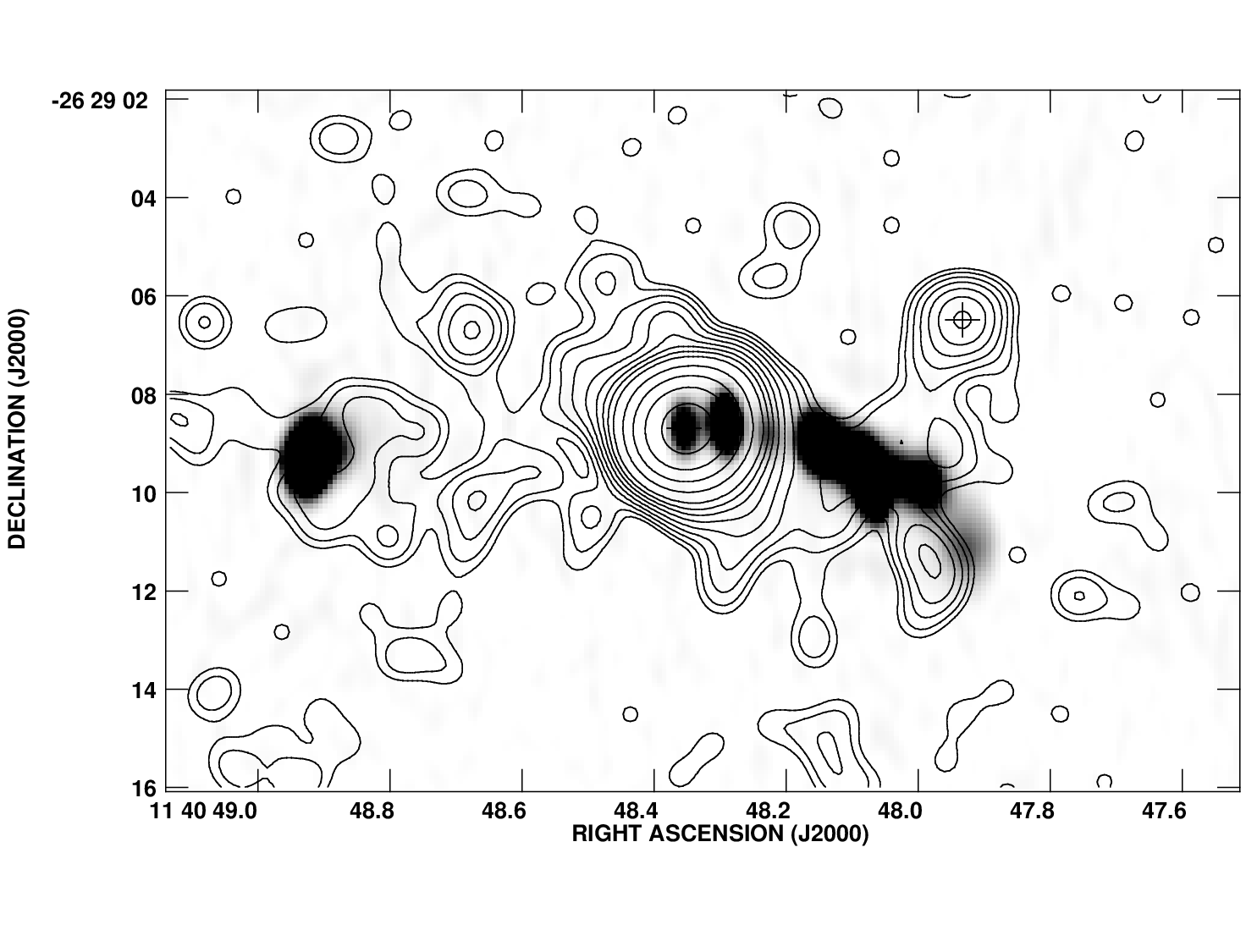,width=4.5in,angle=-90}
\vspace*{-0.5in}
\psfig{figure= 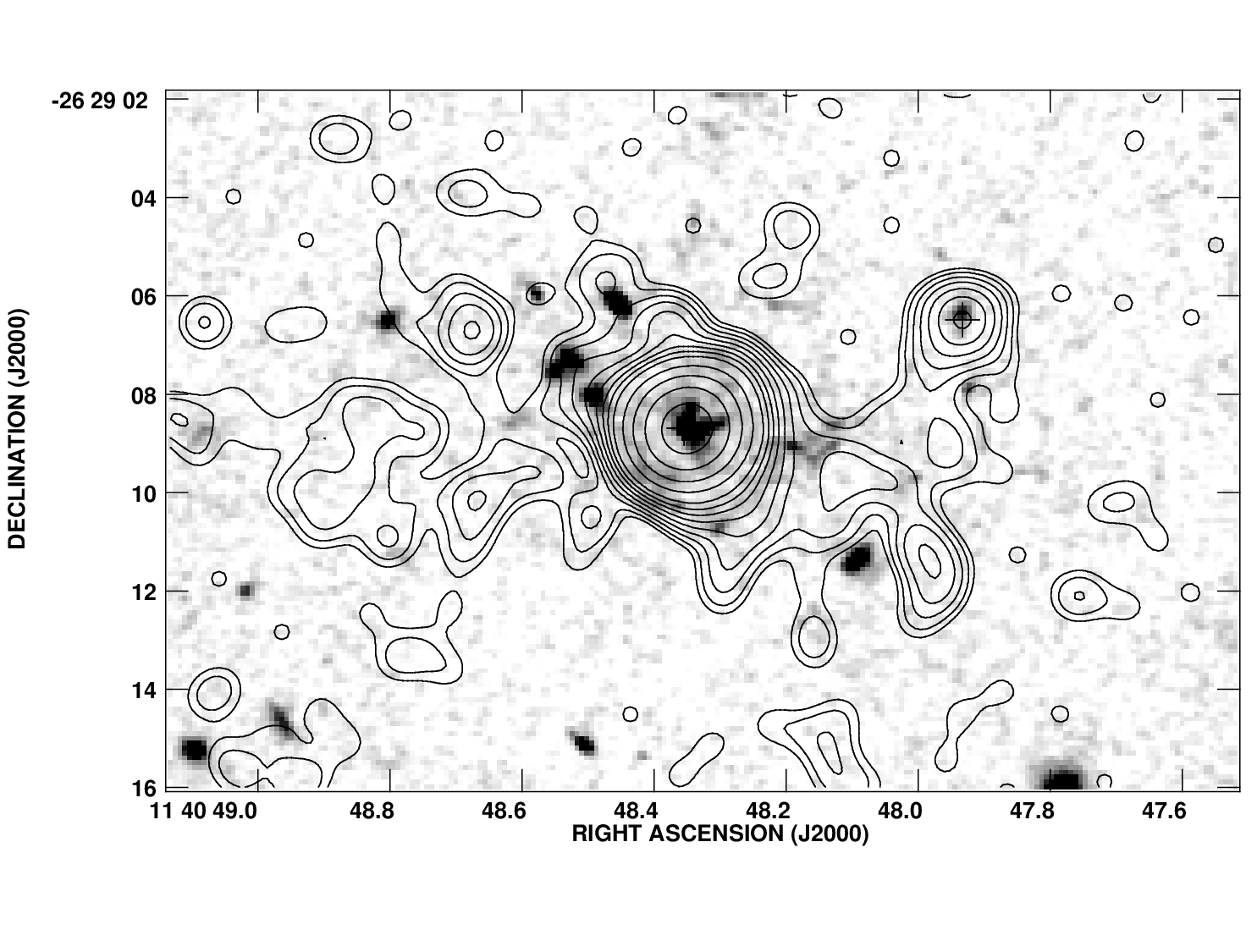,width=4.5in,angle=-90}
\vspace*{-0.5in}
\psfig{figure=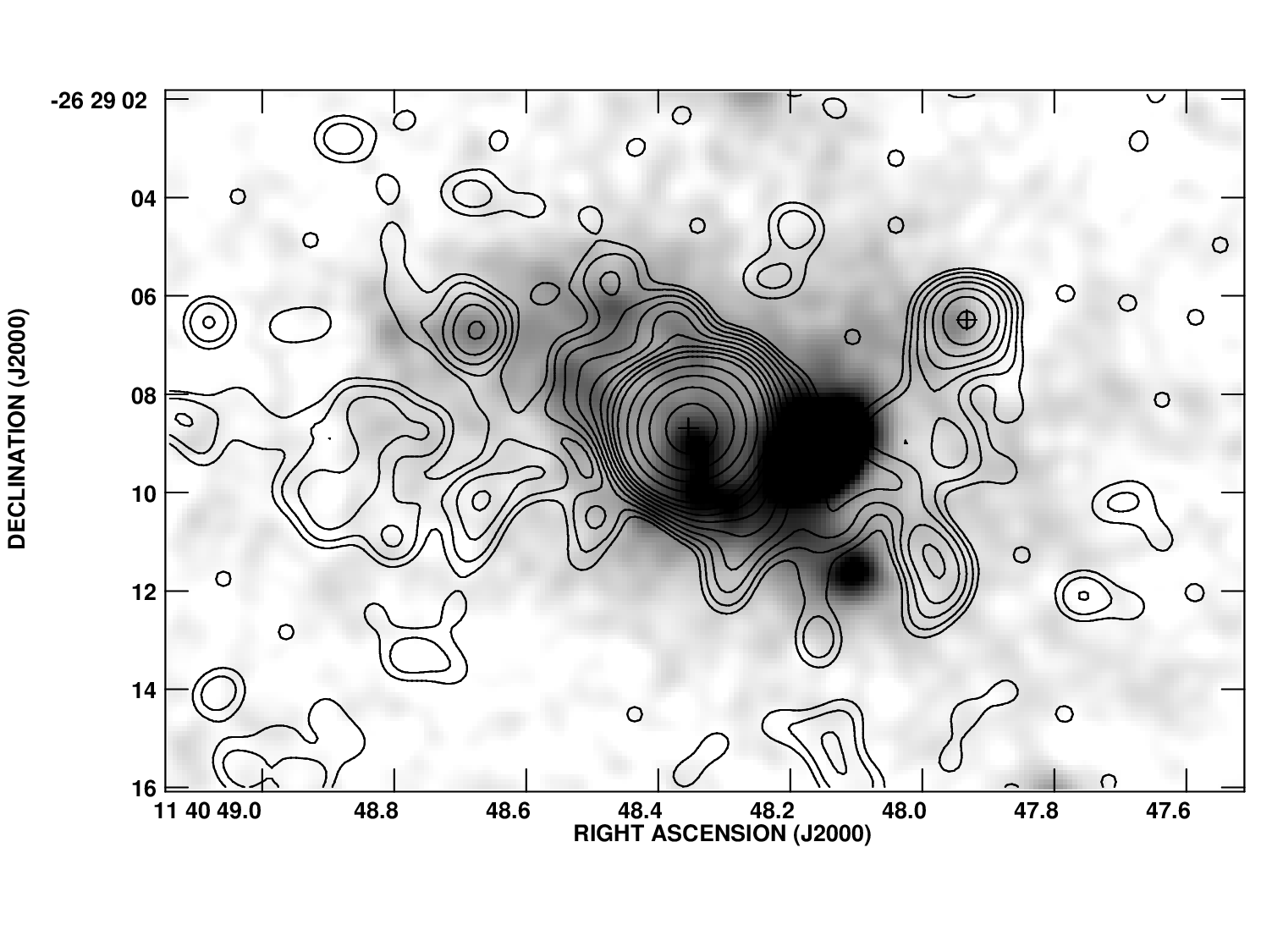,width=4.5in,angle=-90}
\vspace*{-0.4in}
\caption{}
\end{figure}

\clearpage
\newpage

\begin{figure}
\vskip -0.5in
\psfig{figure=1138SOFT.PS,width=4in,angle=-90}
\vspace*{-0.6in}
\psfig{figure=1138MED.PS,width=4in,angle=-90}
\vspace*{-0.6in}
\psfig{figure=1138HARD.PS,width=4in,angle=-90}
\vspace*{-0.4in}
\caption{}
\end{figure}

\clearpage
\newpage

\begin{figure}
\psfig{figure=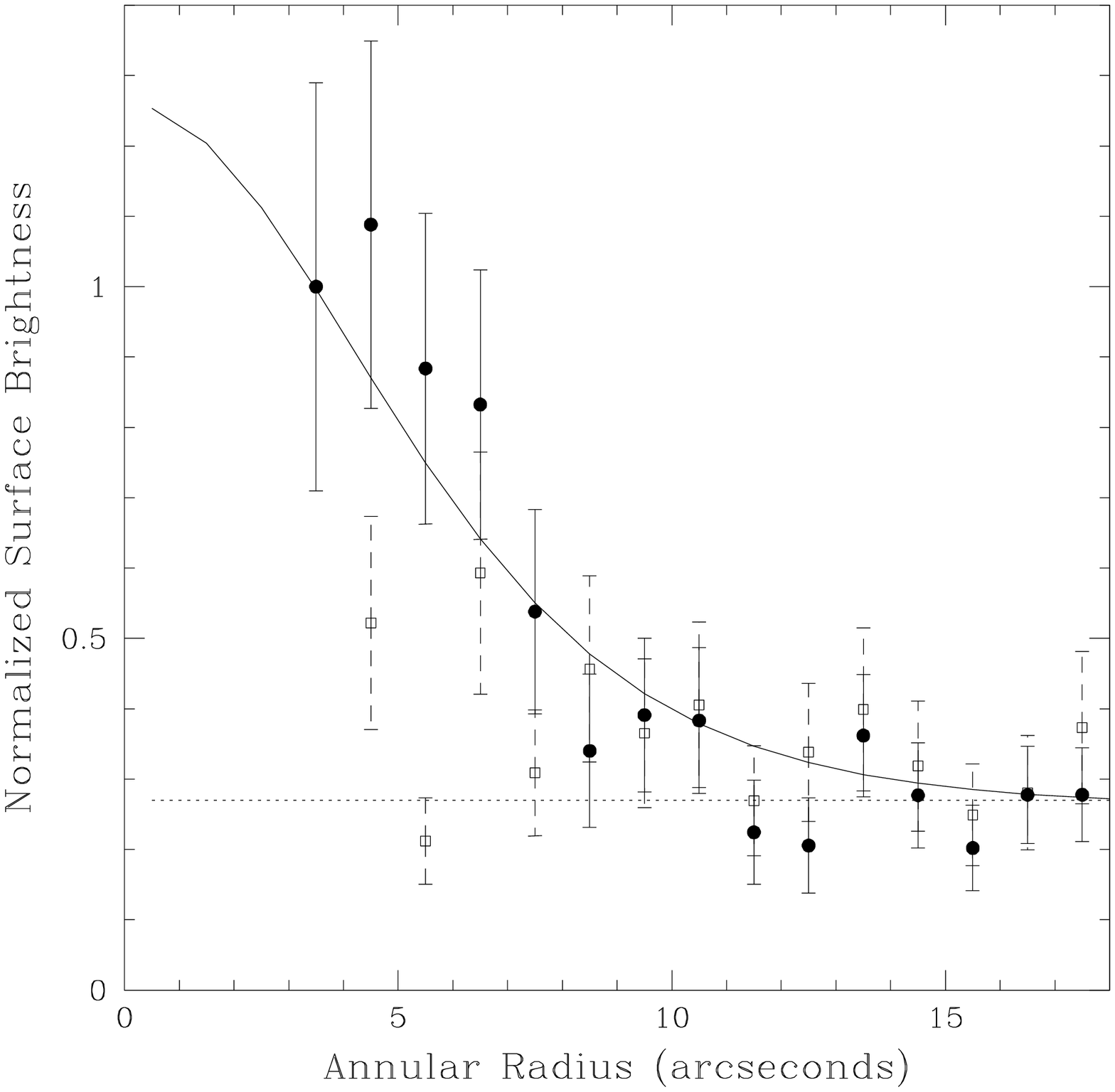,width=6in}
\caption{}
\end{figure}

\clearpage
\newpage

\begin{figure}
\psfig{figure=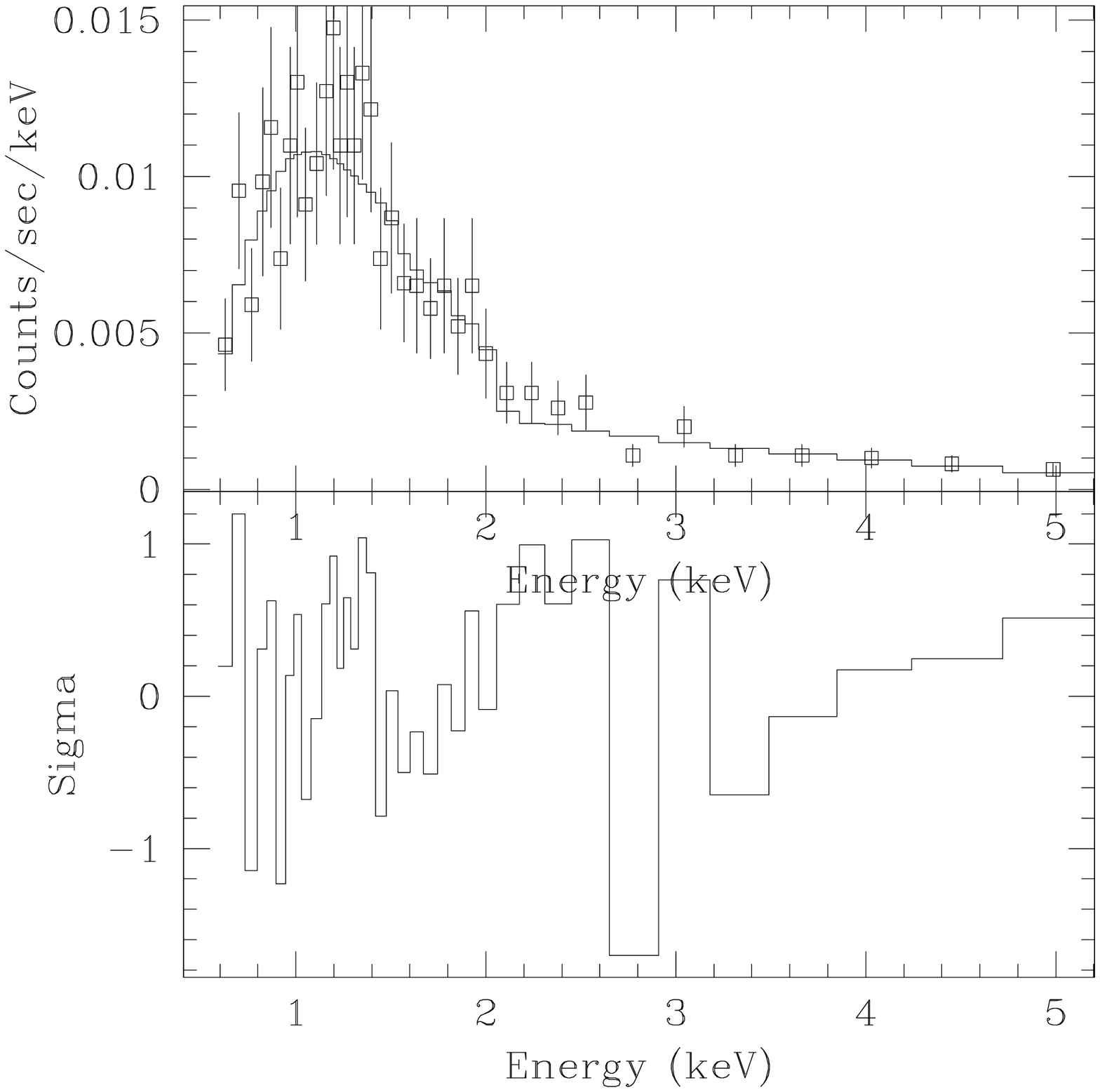,width=6in}
\caption{}
\end{figure}

\end{document}